\documentstyle[epsfig,12pt]{aipproc}
\def \beq{\begin{equation}}
\def \eeq{\end{equation}}
\def \efi{Enrico Fermi Institute Report No.\ EFI}
\begin{document}
\title{Extensive Air Shower Radio Detection:\\
  Recent Results and Outlook
\footnote{Invited talk presented by J. Rosner at RADHEP-2000 Conference, UCLA,
Nov.\ 16--18, 2000, proceedings published by AIP.  \efi~2000-57,
astro-ph/0101089.}}
\author{Jonathan L. Rosner and Denis A. Suprun}
\address{Enrico Fermi Institute and Department of Physics \\
University of Chicago, Chicago, IL 60637 USA}
\maketitle
\begin{abstract}
A prototype system for detecting radio pulses associated with extensive
cosmic ray air showers is described.  Sensitivity is compared
with that in previous experiments, and lessons are noted for future studies.
\end{abstract}

\section{Introduction}

The observation of the radio-frequency (RF) pulse associated with extensive air
showers of cosmic rays has had a long
and checkered history.  In the present report we describe an attempt to
observe such a pulse in conjunction with the Chicago Air Shower Array (CASA)
and Michigan Muon Array (MIA) at Dugway, Utah.  Only upper limits
on a signal have been obtained at present, though we are still processing
data and establishing calibrations.

In Section~2 we review the motivation and history of RF pulse detection.
Section~3 is devoted to the work at CASA/MIA, while Section~4 deals with
some future possibilities, including ones associated with the planned
Pierre Auger observatory.  We conclude in Section~5.  This report is an
abbreviated version of a longer one \cite{nim}.

\section{Motivation and history}

\subsection{Auxiliary information on shower}

Present methods for the detection of an extensive air shower of cosmic rays
leave gaps in our information.  The height above ground at which showers
develop cannot be provided by ground arrays, though stereo detection by air
fluorescence detectors is useful.  Composition of the primary particles,
another unknown, is correlated with shower height, with heavy
primaries leading to showers which begin higher in the atmosphere.

Radio detection can help fill such gaps.  The electric field associated with a
charge $|e|$ undergoing an apparent angular acceleration $\ddot{\theta}$ is
$|{\cal E}| = 1.5 \times 10^{-26} \ddot{\theta}$ V/m, where time is measured in
seconds \cite{Allan,Feyn}.  For typical showers the charges of radiating
particles are expected to be able to act coherently
to give a pulse with maximum frequency component $\nu_{\rm max} ({\rm MHz})
\simeq 10^6/R^2 ({\rm m})$, where $R$ is the distance of closest approach of
the shower axis to the antenna.  Showers originating higher in the atmosphere
are expected to have higher-frequency components.  Thus RF detection may be
able to add information on shower height and primary composition, and to
provide a low-cost auxiliary system in projects such as the
Pierre Auger array \cite{Auger}.

\subsection{Pulse generation mechanisms}

Several possibilities have been discussed for generation of a pulse by
air showers.  Cosmic rays could induce the atmosphere to act as a giant spark
chamber, triggering discharges of the ambient field gradient \cite{RRW}.
Compton scattering and knock-on electrons can give rise to a negative charge
excess of some 10 to 25\% at shower maximum \cite{Ask}.  Separation of positive
and negative charges can occur in the Earth's magnetic field as a result of a
$q {\bf v} \times {\bf B}$ force \cite{KL}.  This last mechanism
is thought to be the dominant one accounting for atmospheric
pulses with frequencies in the 30--100~MHz range
\cite{Allan}, and will be taken as the model for the signal for which the
search was undertaken.  The charge-excess mechanism is probably the major
source of an RF signal in a dense material such as polar ice \cite {ZHS},
but is expected to be less important in the atmosphere.

\subsection{Early measurements}

The first claim for detection of the charge-separation mechanism utilized
narrow-band techniques at 44 and 70~MHz \cite{Jelley,Weekes}.
A Soviet group reported signals at 30~MHz \cite{Sov}, while a University of
Michigan group at the BASJE Cosmic Ray Station on Mt.~Chacaltaya,
Bolivia \cite{Chac} studied pulses in the 40--90~MHz range.

The collaboration of H. R. Allan {\it et al.}\ \cite{Allan} at Haverah Park in
England studied the dependence of signals on primary energy
$E_p$, perpendicular distance $R$ of closest approach of the
shower core, zenith angle $\theta$, and angle $\alpha$ between
the shower axis and the magnetic field vector. Their results
indicated that the electric field strength per unit of frequency,
${\cal E}_\nu$, could be expressed as
\beq \label{eqn:E}
{\cal E}_\nu = s \frac{E_p}{10^{17} {\rm~eV}} \sin \alpha \cos \theta
\exp \left( - \frac{R}{R_0(\nu, \theta)} \right)~~~\mu{\rm
V}~m^{-1}~ {\rm MHz}^{-1}~~~,
\eeq
where $R_0$ is an increasing function of $\theta$, equal (for example) to $(110
\pm 10)$ m for $\nu = 55$~MHz and $\theta < 35^\circ$.  The constant $s$ was
originally claimed to be 20.  The Haverah Park observations were
recalibrated to yield $s = 1.6$ (0.6 $\mu$V m$^{-1}$~MHz$^{-1}$ for a
$10^{17}$~eV shower at $R = 100$ m) while observations in the U.S.S.R. gave
$s = 9.2$ (3.4~$\mu$V~m$^{-1}$~MHz$^{-1}$ at $R = 100$ m) \cite{Atrash}.
To estimate the corresponding signal
strength in \cite{Jelley,Weekes}, we note that the signal power for
showers of average primary energy $E = 5 \times 10^{16}$~eV was measured to be
about 4 times that of galactic noise, for which \cite{Allan} ${\cal E}_\nu^{\rm
Gal} \simeq 1$--$2~\mu$V m$^{-1}$~MHz$^{-1}$.  Thus, for such showers, one
expects ${\cal E}_\nu \simeq 2$--$4~\mu$V m$^{-1}$~MHz$^{-1}$.
Similarly, the estimate \cite{Weekes} of an average pulse power $V_{\rm peak}^2
/2 R = 10^{-12}$ W gives $V_{\rm peak} = 10~\mu$V for $R = 50~\Omega$.  Using
the relation between pulse voltage and ${\cal E}_\nu$ \cite{Allan}
\beq \label{eqn:VE}
V = 30 G^{1/2} (\delta \nu/\nu) {\cal E}_\nu~~~,
\eeq
where
$G$ is the antenna gain, and $\delta \nu$ is the bandwidth centered at
frequency $\nu$, we find for an assumed $G = 5$ (7~dB) (it is not quoted in
Ref.\ \cite{Jelley}) and $\delta \nu/ \nu = 2.75/44$ \cite{Jelley}, a value of
${\cal E}_\nu \simeq 2.4~\mu$V m$^{-1}$~MHz$^{-1}$ at a primary energy of
$5 \times 10^{16}$~eV, or about $5~\mu$V m$^{-1}$~MHz$^{-1}$ at $10^{17}$~eV if
${\cal E}_\nu$ scales linearly with primary energy \cite{Allan}.  For $G=5$ the
data of Refs.\ \cite{Jelley,Weekes} thus would favor the higher field-strength
claims of the U.S.S.R. group cited in Ref.\ \cite{Atrash}.

More recent claims include pulses with
components at or below several MHz \cite{Agasa,Yak,GS,Gau}, and
at VHF frequencies \cite{GS,Gau}.  The Gauhati University group has reviewed
evidence for pulses at a wide range of frequencies \cite{Gau}.

\subsection{Pulse characteristics}

The Haverah Park observations are consistent with a model
in which the pulse's onset is generated
by the start of the shower at an elevation of about 10 km above
sea level, while its end is associated with the greater total
path length (shower $+$ signal propagation distance) associated
with the shower's absorption about 5~km above sea level.  If a vertical
shower is observed at a distance of 100~m from its core, the pulse should
rise and fall back to zero within about 10~ns, with a subsequent longer-lasting
negative component.  High frequencies should be less visible far from
the shower axis.  Heavy primaries should lead to showers originating higher in
the atmosphere, with consequent higher-frequency RF components as a result
of the geometric aspect ratio with which they are viewed by the antenna, and
possibly a greater ${\cal E}_\nu$ for a given primary energy \cite{Allan}.
The polarization of the pulse should be dictated by the mechanism of pulse
generation:  e.g., perpendicular to the line of sight with component
along ${\bf v} \times {\bf B}$ for the charge-separation mechanism.

\subsection{RF backgrounds}

Discharges of atmospheric electricity will be detected at random
intervals at a rate depending on local weather
conditions and ionospheric reflections.  Man-made RF
sources include television and radio stations, police and other
communications services, broad-band sources (such as ignition
noise), and sources within the experiment itself.  The
propagation of distant noise sources to the receiver is a strong
function of frequency and of solar activity.

Galactic noise can be the dominant signal in exceptionally
radio-quiet environments for frequencies in the low VHF (30--100~MHz)
range \cite{Allan}. For higher frequencies in such
environments, thermal receiver noise becomes the dominant effect.

\section{The installation at CASA}

The CASA/MIA detector is located about 100~km southwest of Salt Lake City,
Utah, at the Dugway Proving Ground
\cite{CASAnim}.  The Chicago Air Shower Array (CASA) is
a rectangular grid of $33 \times 33$ stations on on the desert's surface.
The inter-station spacing is 15~m. A station has four 61
cm $\times$ 61~cm $\times$ 1.27~cm sheets of plastic scintillator
each viewed by its own photomultiplier tube (PMT).  When a signal
appears on 3 of 4 PMTs in a station, a ``trigger request pulse''
of 5~mA with 5~$\mu$s duration is sent to a central trailer, where
a decision is made on whether to interrogate all stations for a
possible event.  Details of this trigger are described in Ref.~\cite{CASAnim}.
When this experiment was begun the CASA array had been reconfigured to remove
the 4 westernmost ``ribs'' of the array.  For
runs performed in 1998, the easternmost rib had also been removed.

The University of Michigan designed and built a muon detection
array (MIA) to operate in conjunction with CASA.  It consists of
sixteen ``patches,'' each having 64 muon counters, buried 3~m
below ground at various locations in the CASA array.
Each counter has lateral dimensions 1.9 m $\times$ 1.3
m.  Four of the patches (numbered 1 through 4), each about 45~m
from the center of the array, lie on the corners of a skewed
rectangle; four (numbered 5 through 8), each about 110 m from the
center of the array, lie on a rectangle with slightly different
skewed orientation, and eight (numbered 9 through 16) lie on the
sides and corners of a rectangle with sides $x \simeq \pm 180$~m
and $y \simeq \pm 185$~m, where $x$ and $y$ denote East and North
coordinates.

\subsection{Expected integral rates}

The CASA trigger threshold is a few $\times 10^{14}$~eV and corresponds to
a trigger rate of 10--20 Hz.  The expected rates above ($10^{15},~10^{16},~
10^{17},~10^{18}$)~eV are about (1~Hz, 1 per 2 min, 1 per 4~hr, and 2 per mo),
respectively, over the 1/4 km$^2$ area of the array. 
A primary energy of at least $10^{17}$~eV seems to be
needed if radio signals are to exceed the galactic noise level of
1--2 $\mu$V m$^{-1}$~MHz$^{-1}$.  Such pulses could be generated by a 
$10^{17}$~eV vertical shower with axis 100~m from the antenna under the most
optimistic estimates.  Since the whole array should see such
showers only every few hours, and most have axes farther from the
antenna than 100 m, the possibility of accidental noise pulses during such
long time intervals reduces the expected sensitivity considerably.

\subsection{The ``radio shack'' at CASA}

A survey of the CASA/MIA site determined that within the array, broad-band
noise associated with computers, switching power supplies, and other
electronics was so intense that no RF searches could be undertaken.  The same
was true at any position within the perimeter of the array.
Consequently, an antenna was mounted on top of a mobile searchlight tower at
a height of 10~m about 24~m east of the eastern edge of the array,
corresponding to $x = 263.8$~m,
$y=0$~m.  The antenna, a 9-element portable log-periodic antenna manufactured
by Dorne and Margolin, was acquired from FairRadio Co.~in Lima, Ohio, for about
\$60.  Its nominal bandwidth is 26--76~MHz but it was measured to have usable
properties up to 170~MHz.

The signal was fed through 60~ft.\ of RG-58U cable,
filtered by a high-pass filter admitting frequencies
above 23~MHz, preamplified using a Minicircuits ZFL-500LN preamplifier with
26~dB of gain, low-pass-filtered to admit frequencies below 250~MHz, and fed
to the oscilloscope at a sensitivity of 5~mV/division.  This constituted the
``wide-band'' configuration used for most data acquisition runs.  The filters
were Minicircuits BNC coaxial models.  A ``narrow-band'' configuration with
response between 23 and 37~MHz and two preamplifiers had substantially poorer
signal-to-noise ratio in distinguishing transient signals from background.

A trigger based on the coincidence of seven of the eight outer muon ``patches''
was set to select large showers.  Each muon patch was set to produce a
trigger pulse when at least 5 of its 64 counters registered a minimum-ionizing
pulse within $5.2$~$\mu$s of one another.  The pulses were then combined
to produce a summed pulse, fed to a discriminator, whose output was amplified
and sent over a cable (with measured delay time 2.15~$\mu$s) to the RF trailer.
No evidence for pickup of the trigger pulse from the antenna was found.  The
trigger corresponded to a minimum shower energy somewhat below $10^{16}$~eV,
based on the integral rate \cite{REF} at $10^{18}$~eV of 0.17/km$^2$/day/sr.

A Tektronix TDS-540B digitizing oscilloscope registered filtered
and preamplified RF data on a rolling basis. These data were then captured
upon receipt of a large-event trigger and stored on hard disk using a National
Instruments GPIB interface.  Data were
taken using separate computers (at different times), allowing
for analysis both at the University of Washington and at
Chicago.  The Washington system used a Macintosh Quadra 950
running Labview, while the Chicago system used either a Dell XPS200s Pentium
desktop computer or a Dell Latitude LM laptop computer
running a C program adapted from those provided by National Instruments.
Each trigger caused 50~$\mu$s of RF data, centered around the
trigger and acquired at 1~GSa/s, to be saved.

The total trigger rate ranged between about 20 and 50 events per hour,
depending on intermittent noise sources in the trigger system.
Concurrently, the CASA on-line data acquisition system was
instructed to write files of
events in which at least 7 out of the 8 outermost muon patches
produced a patch pulse.  These files 
typically overlapped with the records taken at the RF trailer to
a good but not perfect extent as a result of occasional noise on
the trigger line.

\subsection{Raw data and RF backgrounds}

To remove strong Fourier components associated with narrow-band RF
signals which were approximately constant over the duration of
each data record, a MATLAB routine performed
the fast Fourier transform of the signal and renormalized the
large Fourier components to a given maximum intensity.  Fig.~1
shows the fast Fourier transform of a typical RF signal before
and after this procedure was applied. In each case the data were
acquired using the ``wide-band'' filter configuration mentioned above, whose
response cuts off sharply below 23~MHz.

\begin{figure}
\centerline{\epsfysize = 4.7 in \epsffile{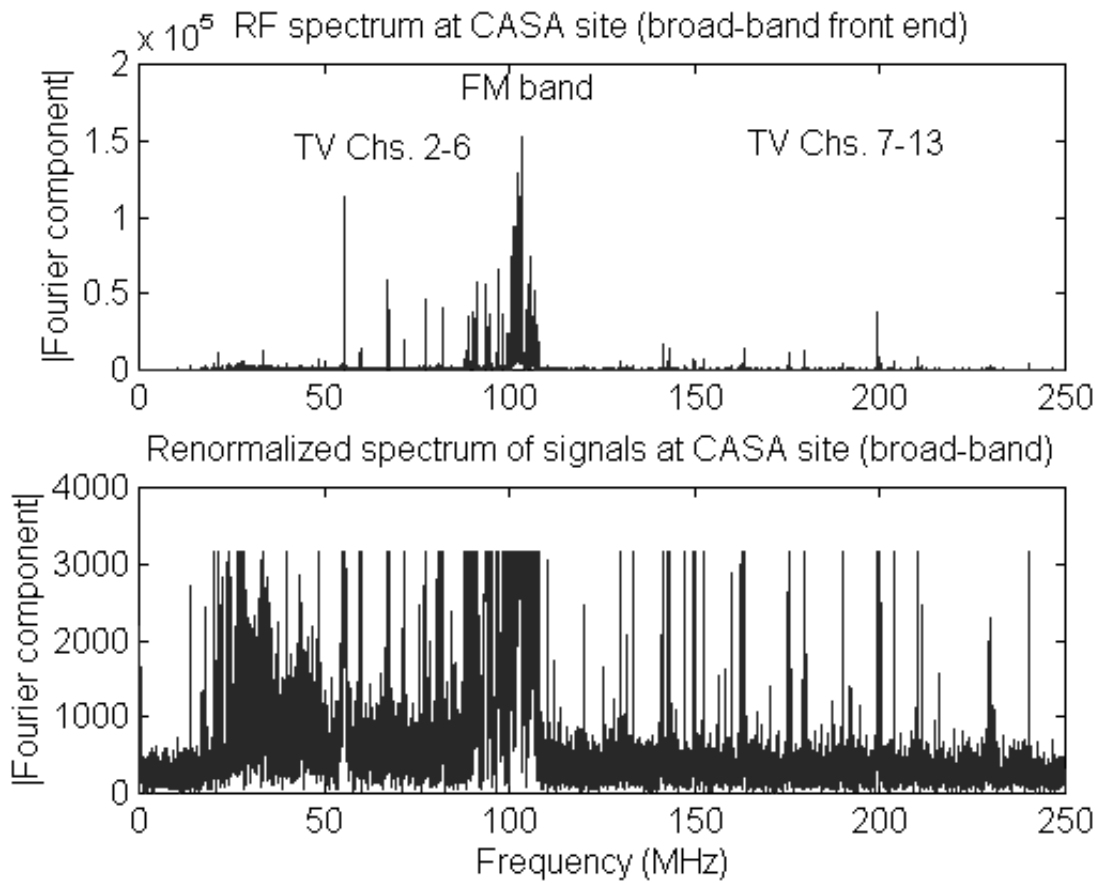}} \caption{Top
panel: Fourier spectrum (in arbitrary units) of RF signals
at Dugway site.  Prominent features include video and audio carriers
for TV Channels 2, 4, 5, 7, and 11,
and the FM broadcast band between 88 and 108~MHz. Bottom panel:
Fourier spectrum (same vertical scale) after renormalization of
large Fourier components to an arbitrary maximum magnitude.  The
continuum between 23 and 88~MHz was not detectable in Chicago;
TV and FM signals were found to be almost 40~dB stronger there,
so gain was reduced correspondingly.}
\end{figure}

\begin{figure}
\centerline{\epsfysize = 7 in \epsffile{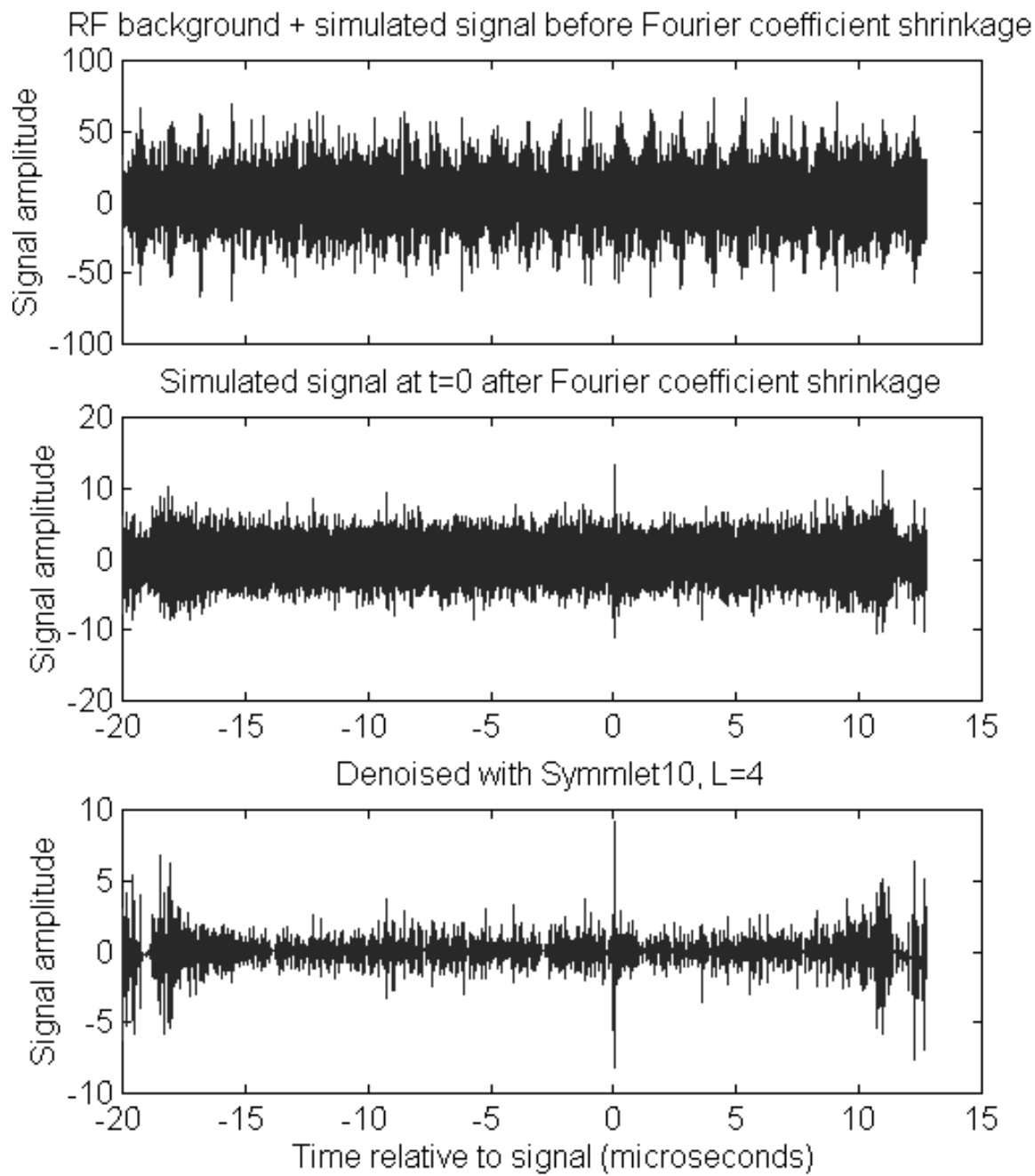}}
\caption{Effect of Fourier coefficient shrinkage on detectability
of a transient.  Top panel:  raw RF record (in arbitrary units)
with simulated signal superposed.  Middle panel:  record (same
scale) after Fourier coefficient shrinkage.  Here a maximum
Fourier coefficient magnitude of $10^3$ (in the units of Fig.~1)
has been imposed.  Bottom panel: the same record after denoising
with a wavelet routine.}
\end{figure}

The effect of digital filtering on detectability of a transient is
illustrated in Fig.~2.  The top panel shows the RF record whose
Fourier transform was given in Fig.~1, on which has been
superposed a simulated transient of peak amplitude 14.5
digitization units. (The data acquisition scale ranges from $-128$ to $+127$
digitization units; one scale division on the oscilloscope corresponds to 25
units.)  The transient is invisible beneath the large
amplitude associated with television and FM radio signals.  The
middle panel shows the result after application of the Fourier
coefficient shrinkage algorithm.  The bottom panel shows the same record
after denoising with a wavelet routine \cite{DFS}.
The records in Figs.~1 and 2 were obtained for 32,768 data points
obtained at a 1~ns sampling interval, with the trigger at the 20,000th point.

\subsection{Signal simulation}

To quantify signal processing efficiency, we generated the expected signal,
fed it through the same preamplifier and filter configurations used for
data acquisition, and superposed it on records otherwise free of
transients.  We successively reduced the amplitude of the
superposed test signal until it was no longer detectable, thereby
obtaining an estimate of sensitivity.

\begin{figure}
\centerline{\epsfysize = 4.7 in \epsffile{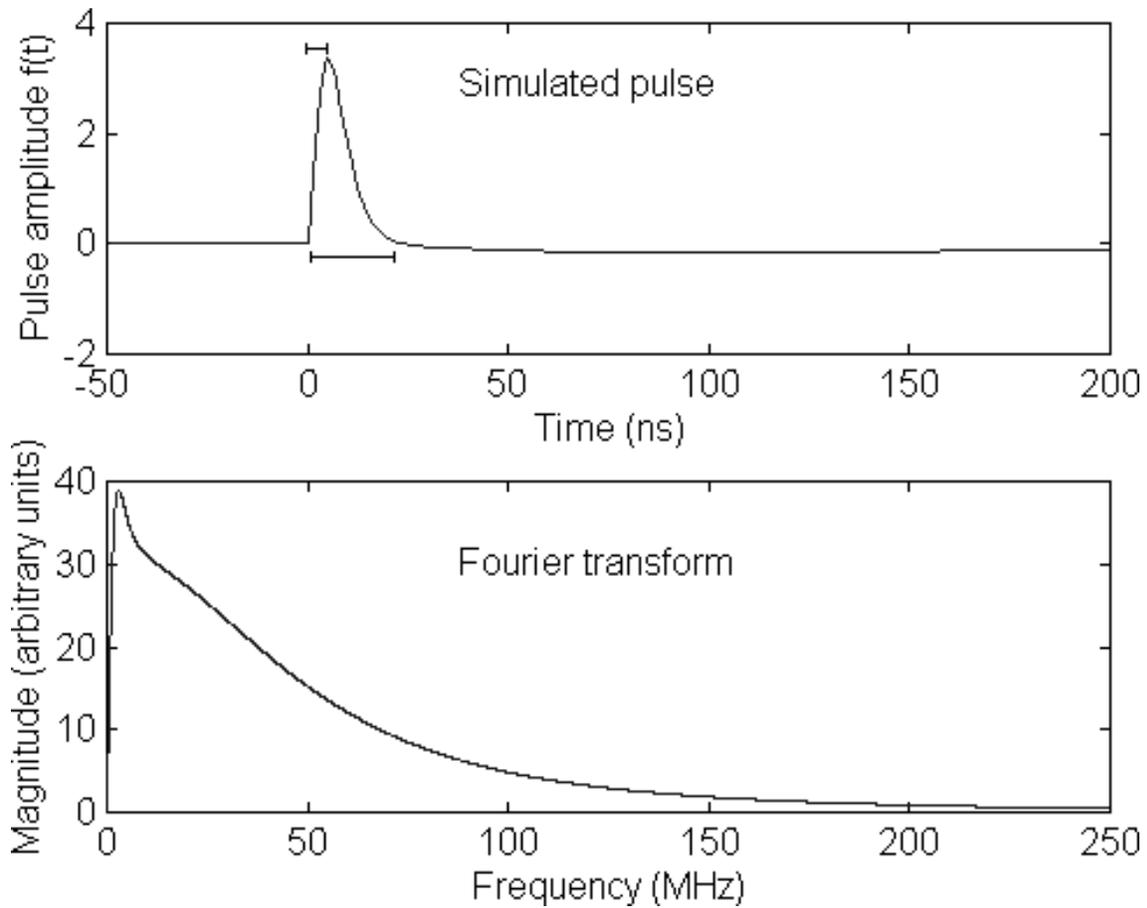}}
\caption{Analytic depiction of typical pulse presented to
filter-preamplifier configuration.  Top panel:  time dependence
of pulse $f(t) = \theta(t) t^2[e^{-0.4t} - e^{-0.02t}/ 8000] (t$
in ns); bottom panel: Fourier spectrum of pulse for $-20~\mu$s
$\le t \le 12.768~\mu$s (calculated analytically). In the top panel, the short
bar above the pulse denotes $\delta$, the time difference between onset and
maximum, while the longer bar below the pulse denotes
the duration of the positive component.}
\end{figure}

A Hewlett-Packard Arbitrary Waveform Generator was used to
generate signals whose typical characteristics are illustrated in Fig.~3.
These signals were taken to have the form $f(t) = \theta(t)
At^2(e^{-Bt} - C e^{-Dt})$ with the coefficient $C$ chosen so
that $f(t)$ has no DC component, and $D$ corresponding to a long
duration of the negative-amplitude component.  For all pulses we
chose $D = B/20$, so that $C = (8000)^{-1}$ cancels the DC
component.  The Fourier components of the test pulse fall off
smoothly with frequency.  The initial $t^2$ behavior was chosen
so that both the test pulse and its first derivative vanish at
$t=0$, as might be expected for a pulse from a developing shower.  We
chose $B = 0.8,~0.4,~0.2,~0.1$ corresponding to a time difference between
pulse onset and maximum of $\delta = 2.5~,5,~10,~20$~ns
and simulated both narrow-band (23--37~MHz) and broad-band (23--250~MHz)
configurations.

The shape of the pulse of Fig.~3 is affected by preamplification and filtration
as shown in Fig.~4 for the broad-band example.  (The narrow-band configuration
leads to a longer ringing time.)  The noise is associated with the system used
to generate the test pulse, and the fact that the Fourier transform
is taken over a much longer time than the duration of the pulse.  The sharp
feature at 125~MHz is a local artifact.

Systematic studies of signal-to-noise ratios have been performed
so far only for the simulated pulses with $\delta=5$~ns applied
to the broad-band front end.  A typical pulse of this type
gave a front end output of 21~mV peak-to-peak, acquired at an
oscilloscope sensitivity of 5~mV per division.  Since each division
corresponds to 25 digitization units, the peak-to-peak range
is about 104 digitization units, or slightly less than half the
dynamic range (255 units, or 8 bits).  Positive and negative
peaks are thus about 52 digitization units each.

The stored test signal is then multiplied by a scale factor and
added algebraically to a collection of RF records in which, in
general, randomly occurring transients will be present.  One then
inspects these records to see if the transient can be
distinguished from random noise.

For the broad-band data we estimated that pulses with input
voltages corresponding to about 1/5 the original test pulse can
be distinguished from average noise (not from noise spikes!).
Since the original test pulse had a peak value of 1.3~mV, this
corresponds to sensitivity to an antenna output of about $V_{\rm
pk} \simeq 260~\mu$V.  The ability to detect such a pulse with an
effective bandwidth of about 30~MHz corresponds to a threshold
sensitivity at the level of order $3~\mu$V/m/MHz \cite{nim}.

\begin{figure}
\centerline{\epsfysize = 4.7 in \epsffile{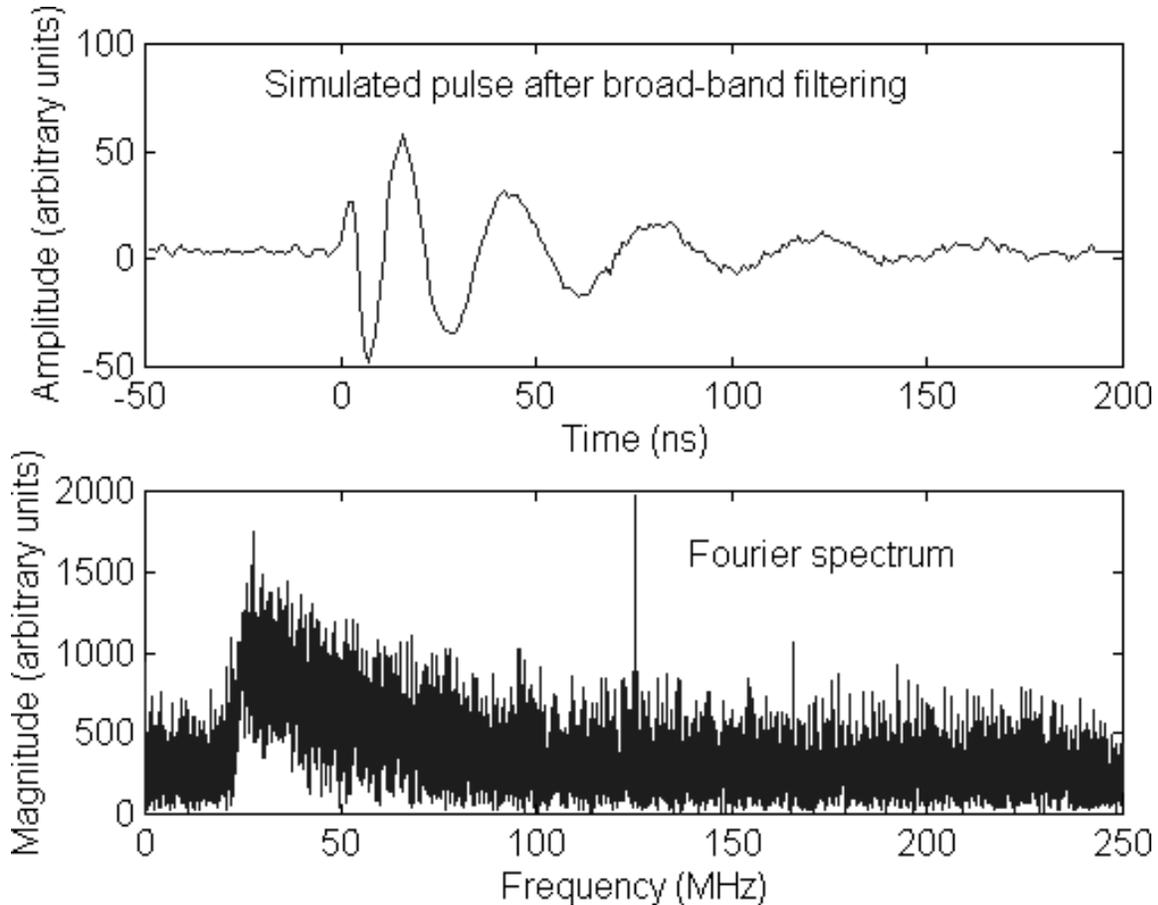}}
\caption{Test pulse of Fig.~3 after broad-band filtration (23--250~MHz) and
preamplification. Top panel:  time dependence of pulse; bottom
panel: Fourier spectrum of recorded pulse for $-20~\mu {\rm s}
\le t \le 12.768~\mu$s.}
\end{figure}

Preliminary studies of simulated pulses applied to the
narrow-band front end suggest a considerably poorer achievable
signal-to-noise ratio, despite the expectation that the signal
should have a large portion of its energy between 23 and 37~MHz.
It appears difficult to detect a pulse from the antenna below
about 0.7~mV, which for a bandwidth of 14~MHz corresponds to a
threshold sensitivity of $7~\mu$V/m/MHz \cite{nim}.
Studies of possible improvements of the analysis
algorithm for the narrow-band data are continuing.

\subsection{Transients detected under various conditions}

\begin{figure}
\centerline{\epsfysize = 5 in \epsffile{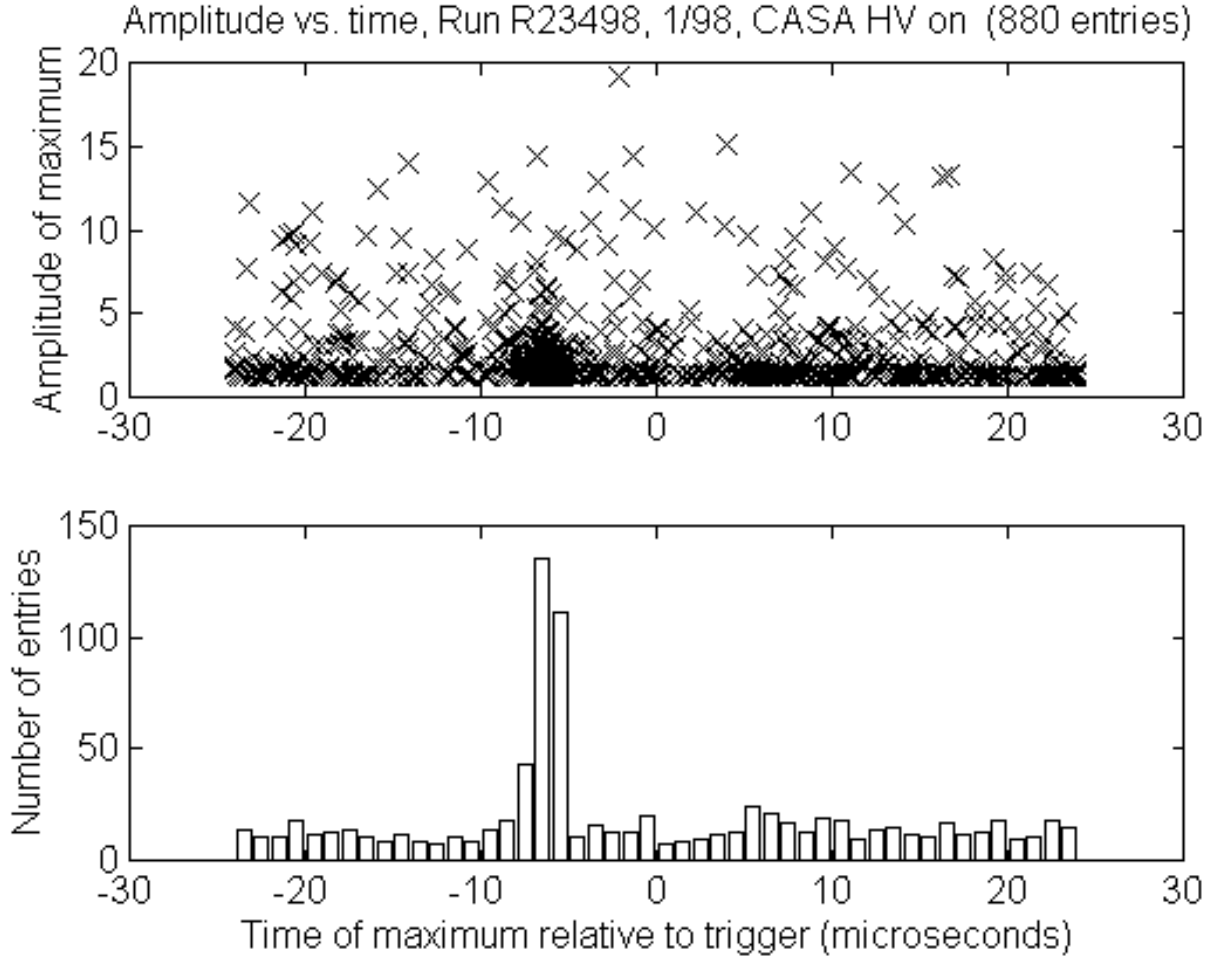}} \caption{Top
panel:  time-vs.-amplitude plot for maxima of 880 events recorded
in January 1998 (RF run 23498 only) with CASA HV supplied to all stations.  All
events recorded with East-West antenna polarization.  Bottom
panel: time distribution of transients.}
\end{figure}

Several means were used to characterize transients.  One method
with good time resolution involved the shrinkage of large Fourier coefficients.
One can then search for peaks of each data record, plotting their
amplitude against time relative to the trigger.  One such
plot is shown in Fig.~5 for a data run in which CASA HV was
delivered to all boxes.  A strong accumulation of transients,
mostly with amplitude just above the arbitrarily chosen threshold
(mean + 3 $\sigma$), is visible at times $-5$ to $-7~\mu$s relative to
the trigger.  In a comparable plot for a run in which CASA HV was
completely disabled (Fig.~6), only a small accumulation at times $-6$ to
$-7~\mu$s is present.  This excess appears due to transients with
predominantly high-frequency components (over 100~MHz).  Since signal
pulses are expected to have more power below 100~MHz (see Fig.~4, bottom)
we believe that this accumulation is not due to shower radiation, but
most likely arises from the muon patches, one of which is within
75~m of the antenna. 

A typical transient occurring in a run with CASA HV on is shown in Fig.~7.  The
transients are highly suppressed (though not in all runs) when
CASA boxes within 100 m of the antenna are disabled.

The time distribution of event maxima above an arbitrary
threshold for 880 events taken with CASA HV on (one run
from January 1998) is shown in the bottom panel of Fig.~5.  The
mean arrival time is about $6~\mu$s before the trigger, with a
distribution which is slightly broader for pulses arriving
earlier than the mean.  This broadening may correspond to some
jitter in forming the trigger pulse from the sum of muon patch
pulses.

\begin{figure}
\centerline{\epsfysize = 5 in \epsffile{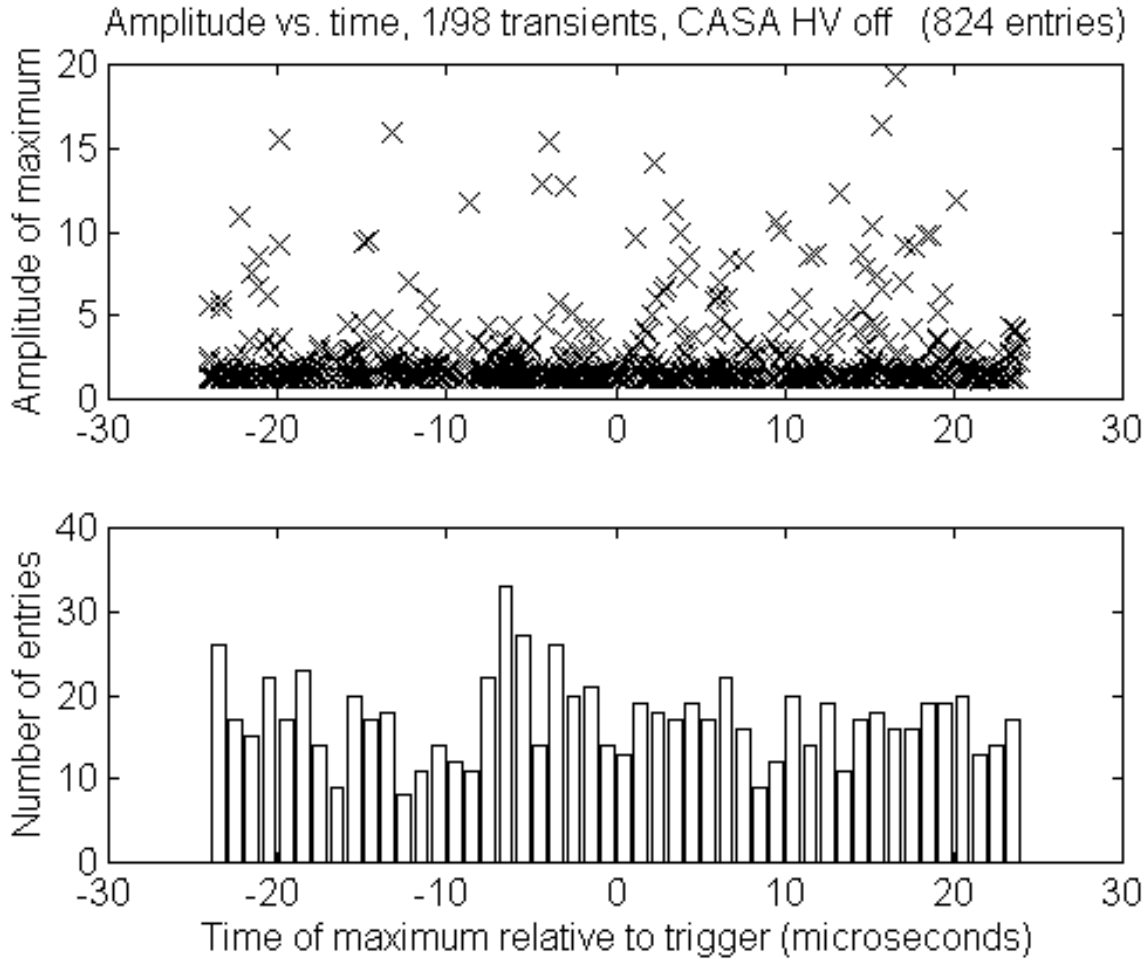}} \caption{Top
panel: time-vs.-amplitude plot for maxima of 824 events recorded
in January 1998 with CASA HV disabled.  All events recorded with
East-West antenna polarization.  Bottom panel:  time distribution
of transients.}
\end{figure}

As mentioned earlier, the time for the trigger pulse to propagate
from the central station to the RF trailer was measured to be
$2.15~\mu$s.  One expects a similar travel time for pulses to
arrive from muon patches to the central station.
Moreover, the muon patch signals are subjected to delays so that
they all arrive at the central station at the same time for a
vertically incident shower.  Thus, the peak in Fig.~5 is
consistent with being associated with the initial detection of a
shower by CASA boxes.  This circumstance was checked by recording
CASA trigger request signals simultaneously with other data; they
coincide with transients such as those illustrated in Fig.~7
within better than $1/2~\mu$s.

\begin{figure}
\centerline{\epsfysize = 4.9 in \epsffile{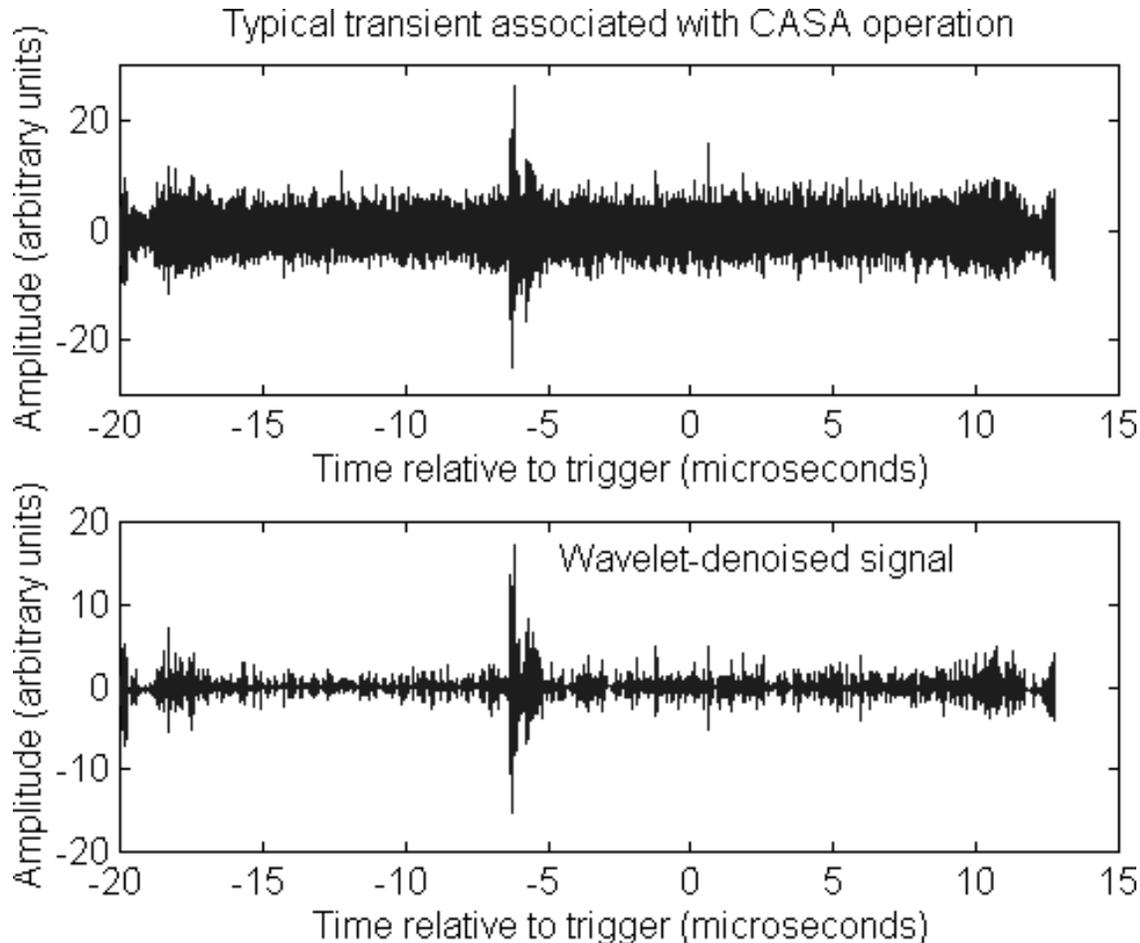}}
\caption{Signal of a typical transient associated with CASA
operation.  Top panel:  before denoising; bottom panel:  after
denoising.}
\end{figure}

The RF signals from the shower are expected to arrive around the
same time as, or at most several hundred nanoseconds before, the
transients associated with CASA operation.  They would propagate
directly from the shower to the antenna, whereas transients from
CASA stations are associated with a slightly longer total path
length from the shower via the CASA station to the antenna. There
will also be some small delay at a CASA station in forming the
trigger request pulse.  Thus, we expect a genuine signal also to
show up around 6--7 $\mu$s before the trigger. However, for data
recorded with CASA boxes disabled, no significant peak with the
expected frequency spectrum is visible
in this time window. The upper limit on the rate of events giving
rise to such a peak can be used to set a limit on RF
pulses associated with air showers.

In Table \ref{tab:triggs} we summarize the triggers taken under optimum
conditions, corresponding to broad-band data acquisition under conditions
of minimum ambient noise.  These triggers represent about 50 hours of data.
Data were taken with both East-West and North-South antenna polarizations.
Since noise from CASA boxes was found to be a significant source of RF
transients, data were taken with some or all CASA boxes disabled
by turning off high voltage (HV) supply to the photomultipliers.

\begin{table}
\caption{Broad-band data recorded under lowest-noise conditions.
\label{tab:triggs}}
\begin{center}
\begin{tabular}{l c c c c} \hline
Antenna      & CASA  &  CASA  & Partial & Total  \\
Polarization & HV on & HV off & CASA HV & events \\ \hline
East-West    &  4503 &   857  &  1834   &  7194  \\
North-South  &   677 &   641  &   366   &  1821  \\ \hline Total
events &  5180 &  1498  &  2200   &  9015  \\ \hline
\end{tabular}
\end{center}
\end{table}

\subsection{Sensitivity estimate}

The main difficulty associated with pulse detection is that signal pulses are
not easily distinguishable from large spurious pulses originating from
atmospheric discharges.  Both air shower pulses and these background noise
pulses can considerably exceed the average noise level.  We estimate that all
signal pulses should arrive between 7 and 6 $\mu$s before the trigger, while
the time distribution of noise pulses is assumed to be uniform.  Hence, a
sufficiently large relative accumulation of pulse maxima in that time bin was
adopted as a key criterion in search of signal pulses. We also employed several 
criteria to increase the fraction of signal pulses: The pulse should be larger 
than some specified magnitude threshold, bandwidth limited and approximately 
uniform within its limited bandwidth. To date, no significant accumulation has 
been detected.

At present we are only able, after accounting for the rate of accidental
noise pulses, using Monte-Carlo simulation, and taking an antenna gain of
$G = 2.5$ and a cable attenuation factor of 1.4~dB, to set an upper limit of 
$s = 54$ in Eq.~(\ref{eqn:E}).  This is to be compared with the original 
Haverah Park result $s=20$ \cite{Allan}, the recalibrated result $s=1.6$ 
\cite{Atrash}, and the Soviet group's result $s=9.2$ \cite{Atrash}. 
The noise level at Dugway is too high and the acquired sample is too limited
in statistics and dynamic range to allow us to place upper limits strict enough
to check the claims of the two groups. We hope that further
improvement of the data processing technique will reduce the
noise contribution. The magnetic dip angle $\gamma$ is much smaller at the
Auger site in Argentina ($34^{\circ}$ versus $68^{\circ}$ at Utah), leading
one to expect bigger electric fields for vertical showers
and facilitating the detection of shower radiation. We also expect
that the Argentina site will be quieter than Dugway and some
clarity as regards the calibrating factor will be established.

\section{Outlook}

\subsection{Further possibilities for processing present data}

We are still hoping to improve our sensitivity to the point that we can
see a true RF signal from a shower even when CASA HV is not disabled.  Such a
signal should precede CASA-related transients by
at least the delay of formation of a phototube signal.
An important calibration will be achieved if we can determine whether we are
sensitive to galactic noise, which may be responsible for the continuum
between 23 and 88~MHz visible in Fig.~1.

We are still exploring improved methods for removing constant RF signals
from our records.  In this respect we are limited by the 8-bit dynamic range
of the TDS-540B oscilloscope.  Data taken with various configurations of
boxes near the antenna disabled may help us to better characterize the
CASA-related transients.

The triggered data may be useful in a rather different context.  It has
been proposed that radar methods be used to detect ion trails associated
with extensive air showers \cite{PG}.  In our case we may be able to
investigate sudden enhancements of the signals of distant television
signals (on Channels 3, 6, 8, and 12) correlated with receipt of a large-event
trigger.

\subsection{Considerations for Auger site}

A number of questions have been suggested by the present investigation if
RF pulse detection is to be considered as an adjunct to the Auger array.

\begin{itemize}

\item How far from the shower axis can antennas detect pulses from showers with
energies above $10^{18}$ or $10^{19}$~eV?  The answer determines whether a
sparse array (e.g., one with the same density as Auger stations)
would be sensitive to the RF pulse.

\item What dynamic range for a data-acquisition system is needed so that a
transient signal survives digital filtering?  Apparently 8-bit range is not
enough.

\item What RF interference exists at each site?  Surveys would be desirable.
They could pinpoint not only narrow-band sources due to broadcasting stations
but potentially dangerous broadband sources from switching power supplies,
computers, etc.  It would be best to undertake such studies after prototype
systems are in place.

\item What power budget would an RF detection system need?  Each Auger
solar-powered station is limited to a total budget of 10 watts.  Presumably
an RF system would have to use auxiliary power, particularly for its
fast-digitization and memory components.

\item What is the envisioned minimum cost per RF station?  It would presumably
be dominated by the data-acquisition system; the antennae and preamplifiers
would probably be cheap by comparison.  A preliminary estimate is less than
\$3K per station \cite{nim}.

\end{itemize}

The Southern Hemisphere Auger site is progressing well toward an engineering
array of 40 stations, as we have heard at this conference \cite{Zas}.  It is
hoped that in a couple of years an investigation at that site of the
feasibility of RF pulse detection can be undertaken.

\section{Conclusions}

No ``golden signal'' has been seen for an RF transient associated with
extensive air showers of cosmic rays at the CASA site.  With further
processing, the data may permit the setting of useful upper limits on
signals relevant to at least some of the previous claims.  A
number of useful lessons have been learned if a similar technique is to
be tried in conjunction with the Auger project.

\section{Acknowledgements}

It is a pleasure to thank Mike Cassidy, Jim Cronin, Brian Fick, Lucy Fortson,
Joe Fowler, Rachel Gall, Kevin Green, Brian Newport, Rene Ong, Scott Oser,
Daniel F. Sullivan, Fritz Toevs, Kort Travis, Augustine Urbas, and John
Wilkerson for collaboration and support on various aspects of this experiment.
Thanks are also due to Bruce Allen, Dave Besson, Maurice Givens, Peter Gorham,
Kenny Gross,
Dick Gustafson, Gerard Jendraszkiewicz, Larry Jones, Dave Peterson, John
Ralston, Leslie Rosenberg, David Saltzberg,
Dave Smith, M. Teshima, and Stephan Wegerich for
useful discussions.  This work was supported in part by the Enrico Fermi
Institute, the Louis Block Fund, and the Physics Department of the University
of Chicago and in part by the U. S. Department of Energy under
Grant No.~DE FG02 90ER40560.

\def \ajp#1#2#3{Am.\ J. Phys.\ {\bf#1}, #2 (#3)}
\def \apny#1#2#3{Ann.\ Phys.\ (N.Y.) {\bf#1}, #2 (#3)}
\def \app#1#2#3{Acta Phys.\ Polonica {\bf#1}, #2 (#3)}
\def \arnps#1#2#3{Ann.\ Rev.\ Nucl.\ Part.\ Sci.\ {\bf#1}, #2 (#3)}
\def \cmts#1#2#3{Comments on Nucl.\ Part.\ Phys.\ {\bf#1}, #2 (#3)}
\def \cn{Collaboration}
\def \cp89{{\it CP Violation,} edited by C. Jarlskog (World Scientific,
Singapore, 1989)}
\def \epjc#1#2#3{Eur.\ Phys.\ J. C {\bf#1}, #2 (#3)}
\def \f79{{\it Proceedings of the 1979 International Symposium on Lepton and
Photon Interactions at High Energies,} Fermilab, August 23-29, 1979, ed. by
T. B. W. Kirk and H. D. I. Abarbanel (Fermi National Accelerator Laboratory,
Batavia, IL, 1979}
\def \hb87{{\it Proceeding of the 1987 International Symposium on Lepton and
Photon Interactions at High Energies,} Hamburg, 1987, ed. by W. Bartel
and R. R\"uckl (Nucl.\ Phys.\ B, Proc.\ Suppl.\, vol. 3) (North-Holland,
Amsterdam, 1988)}
\def \ib{{\it ibid.}~}
\def \ibj#1#2#3{~{\bf#1}, #2 (#3)}
\def \ichep72{{\it Proceedings of the XVI International Conference on High
Energy Physics}, Chicago and Batavia, Illinois, Sept. 6 -- 13, 1972,
edited by J. D. Jackson, A. Roberts, and R. Donaldson (Fermilab, Batavia,
IL, 1972)}
\def \ijmpa#1#2#3{Int.\ J.\ Mod.\ Phys.\ A {\bf#1}, #2 (#3)}
\def \ite{{\it et al.}}
\def \jhep#1#2#3{JHEP {\bf#1}, #2 (#3)}
\def \jpb#1#2#3{J.\ Phys.\ B {\bf#1}, #2 (#3)}
\def \lg{{\it Proceedings of the XIXth International Symposium on
Lepton and Photon Interactions,} Stanford, California, August 9--14 1999,
edited by J. Jaros and M. Peskin (World Scientific, Singapore, 2000)}
\def \lkl87{{\it Selected Topics in Electroweak Interactions} (Proceedings of
the Second Lake Louise Institute on New Frontiers in Particle Physics, 15 --
21 February, 1987), edited by J. M. Cameron \ite~(World Scientific, Singapore,
1987)}
\def \kdvs#1#2#3{{Kong.~Danske Vid.~Selsk., Matt-fys.~Medd.} {\bf #1}, No.~#2
(#3)}
\def \ky85{{\it Proceedings of the International Symposium on Lepton and
Photon Interactions at High Energy,} Kyoto, Aug.~19-24, 1985, edited by M.
Konuma and K. Takahashi (Kyoto Univ., Kyoto, 1985)}
\def \mpla#1#2#3{Mod.\ Phys.\ Lett.\ A {\bf#1}, #2 (#3)}
\def \nat#1#2#3{Nature {\bf#1}, #2 (#3)}
\def \nc#1#2#3{Nuovo Cim.\ {\bf#1}, #2 (#3)}
\def \nima#1#2#3{Nucl.\ Instr.\ Meth.\ A {\bf#1}, #2 (#3)}
\def \np#1#2#3{Nucl.\ Phys.\ {\bf#1}, #2 (#3)}
\def \npbps#1#2#3{Nucl.\ Phys.\ B Proc.\ Suppl.\ {\bf#1}, #2 (#3)}
\def \PDG{Particle Data Group, L. Montanet \ite, \prd{50}{1174}{1994}}
\def \pisma#1#2#3#4{Pis'ma Zh.\ Eksp.\ Teor.\ Fiz.\ {\bf#1}, #2 (#3) [JETP
Lett.\ {\bf#1}, #4 (#3)]}
\def \pl#1#2#3{Phys.\ Lett.\ {\bf#1}, #2 (#3)}
\def \pla#1#2#3{Phys.\ Lett.\ A {\bf#1}, #2 (#3)}
\def \plb#1#2#3{Phys.\ Lett.\ B {\bf#1}, #2 (#3)}
\def \pr#1#2#3{Phys.\ Rev.\ {\bf#1}, #2 (#3)}
\def \prc#1#2#3{Phys.\ Rev.\ C {\bf#1}, #2 (#3)}
\def \prd#1#2#3{Phys.\ Rev.\ D {\bf#1}, #2 (#3)}
\def \prl#1#2#3{Phys.\ Rev.\ Lett.\ {\bf#1}, #2 (#3)}
\def \prp#1#2#3{Phys.\ Rep.\ {\bf#1}, #2 (#3)}
\def \ptp#1#2#3{Prog.\ Theor.\ Phys.\ {\bf#1}, #2 (#3)}
\def \rmp#1#2#3{Rev.\ Mod.\ Phys.\ {\bf#1}, #2 (#3)}
\def \rp#1{~~~~~\ldots\ldots{\rm rp~}{#1}~~~~~}
\def \si90{25th International Conference on High Energy Physics, Singapore,
Aug. 2-8, 1990}
\def \slc87{{\it Proceedings of the Salt Lake City Meeting} (Division of
Particles and Fields, American Physical Society, Salt Lake City, Utah, 1987),
ed. by C. DeTar and J. S. Ball (World Scientific, Singapore, 1987)}
\def \slac89{{\it Proceedings of the XIVth International Symposium on
Lepton and Photon Interactions,} Stanford, California, 1989, edited by M.
Riordan (World Scientific, Singapore, 1990)}
\def \smass82{{\it Proceedings of the 1982 DPF Summer Study on Elementary
Particle Physics and Future Facilities}, Snowmass, Colorado, edited by R.
Donaldson, R. Gustafson, and F. Paige (World Scientific, Singapore, 1982)}
\def \smass90{{\it Research Directions for the Decade} (Proceedings of the
1990 Summer Study on High Energy Physics, June 25--July 13, Snowmass, Colorado),
edited by E. L. Berger (World Scientific, Singapore, 1992)}
\def \tasi{{\it Testing the Standard Model} (Proceedings of the 1990
Theoretical Advanced Study Institute in Elementary Particle Physics, Boulder,
Colorado, 3--27 June, 1990), edited by M. Cveti\v{c} and P. Langacker
(World Scientific, Singapore, 1991)}
\def \yaf#1#2#3#4{Yad.\ Fiz.\ {\bf#1}, #2 (#3) [Sov.\ J. Nucl.\ Phys.\
 {\bf #1}, #4 (#3)]}
\def \zhetf#1#2#3#4#5#6{Zh.\ Eksp.\ Teor.\ Fiz.\ {\bf #1}, #2 (#3) [Sov.\
Phys.\ - JETP {\bf #4}, #5 (#6)]}
\def \zpc#1#2#3{Zeit.\ Phys.\ C {\bf#1}, #2 (#3)}
\def \zpd#1#2#3{Zeit.\ Phys.\ D {\bf#1}, #2 (#3)}


\begin{thebibliography}{99}

\bibitem{nim}  K. Green, J. L. Rosner, D. A. Suprun, and J. F. Wilkerson,
\nima{498}{256}{2003}.

\bibitem{Allan} H. R. Allan, in {\it Progress in Elementary Particles and
Cosmic Ray Physics}, v. 10, edited by J. G. Wilson and S. G.
Wouthuysen (North-Holland, Amsterdam, 1971), p. 171, and
references therein.

\bibitem{Feyn} R. P. Feynman, R. B. Leighton, and M. Sands, {\it The Feynman
Lectures in Physics,} Addison-Wesley, Reading, Mass., 1963, Sec.\ I-28.

\bibitem{Auger} J. W. Cronin, \rmp{71}{S165}{1998}; \npbps{80}{33}{2000};
D. Zavrtanik, \npbps{85}{324}
{2000}.  For the Pierre Auger Project Design Report see
{\tt http://www.ses-ng.si/public/pao/design.html}.

\bibitem{RRW} R. R. Wilson, Phys.~Rev. {\bf 108}, 155 (1967).

\bibitem{Ask} G. A. Askar'yan, Zh.~Eksp.~Teor.~Fiz.~{\bf 41}, 616 (1961)
[Sov.~Phys.--JETP {\bf 14}, 441 (1962)];
Zh.~Eksp.~Teor.~Fiz.~{\bf 48}, 988 (1965) [Sov.~Phys.--JETP {\bf
21}, 658 (1965)].

\bibitem{KL} F. D. Kahn and I. Lerche, Proc.~Roy.~Soc.~{\bf A 289}, 206
(1966).

\bibitem{ZHS} E. Zas, F. Halzen, and T. Stanev, Phys.~Rev.~D {\bf 45}, 362
(1992).

\bibitem{Jelley} J. V. Jelley {\it et al.}, Nature {\bf 205}, 327 (1965);
Nuovo Cimento {\bf A46}, 649 (1966); N. A. Porter {\it et al.},
Phys.~Lett.~{\bf 19}, 415 (1965).

\bibitem{Weekes} T. Weekes, this conference.

\bibitem{Sov} S. N. Vernov {\it et al.}, Pis'ma v ZhETF {\bf 5}, 157 (1967)
[Sov.~Phys.--JETP Letters {\bf 5}, 126 (1967)];
Can.~J.~Phys.~{\bf 46}, S241 (1968).

\bibitem{Chac} P. R. Barker, W. E. Hazen, and A. Z. Hendel, Phys.~Rev.~Lett.
{\bf 18}, 51 (1967); W. E. Hazen, {\it et al.}, {\it ibid.} {\bf
22}, 35 (1969); {\bf 24}, 476 (1970).

\bibitem{Atrash} V. B. Atrashkevich et al., Yad.~Fiz.~{\bf 28}, 366 (1978).

\bibitem{CASAnim} A. Borione {\it et al.}, Nucl.~Instrum.~Meth.~A {\bf 346},
329 (1994).

\bibitem{Agasa} K. Kadota {\it et al.}, Proc.~23rd International Conference on
Cosmic Rays (ICRC-23), Calgary, 1993, v.~4, p.~262; Tokyo
Workshop on Techniques for the Study of Extremely High Energy
Cosmic Rays, Tanashi, Tokyo, 27 -- 30 Sept. 1993.

\bibitem{Yak} P. I. Golubnichii, A. D. Filonenko, and V. I. Yakovlev,
Izv. Akad. Nauk {\bf 58}, 45 (1994).

\bibitem{GS} C. Castagnoli {\it et al.}, Proc.~ICRC-23, Calgary, 1993,
v.~4, p.~258.

\bibitem{Gau} R. Baishya {\it et al.}, Proc.~ICRC-23, Calgary, 1993, V.~4,
p.~266; Gauhati University Collaboration, paper submitted to this conference.

\bibitem{REF} M.~A.~Lawrence, R. J. O. Reid, and A. A. Watson,
J.~Phys.~G {\bf 17}, 733 (1991).
\bibitem{mutrig} R. Gall and K. D. Green, UMC-CASA note, Aug.~23, 1996
(unpublished).

\bibitem{DFS} D. F. Sullivan, Master's Thesis, University of Chicago, 1999
(unpublished).

\bibitem{PG} P. W. Gorham, ``On the possibility of radar echo detection of
ultra-high energy cosmic ray- and neutrino-induced extensive air
showers,'' hep-ex/0001041, January, 2000 (unpublished).

\bibitem{Zas} E. Zas, this conference.

\end{thebibliography}
\end{document}